\documentstyle[12pt]{article}

\catcode`\@=11
\def\numberbysection{\@addtoreset{equation}{section}
        \def\theequation{\thesection.\arabic{equation}}}
\def\beq{\begin{equation}}
\def\eeq{\end{equation}}
\begin{document}
\begin{titlepage}
\begin{center}
\hfill CERN-TH-6248/91\\
\vskip 1.in
{\large \bf Classical Scattering in $2 + 1$ Gravity with N Spinning Sources}
\vskip 0.5in
Andrea CAPPELLI\footnote{On leave of absence from INFN, Sezione di Firenze.}
\\[.2in]
{\em Theory Division, CERN, Geneva, Switzerland}
\\
\vskip 0.5in
Marcello CIAFALONI and Paolo VALTANCOLI \\[.2in]
{\em Dipartimento di Fisica, Universit\`a di Firenze\\
and INFN, Sezione di Firenze, Italy}
\end{center}
\vskip .5in
\begin{abstract}
The classical dynamics of N spinning point sources in $2+1$
Einstein-Cartan gravity is considered.
It corresponds to the $ISO(2,1)$ Chern-Simons theory, in which
the torsion source is restricted to its intrinsic spin part.
A class of explicit solutions is found for the dreibein and the spin
connection, which are torsionless in the spinless limit.
By using the residual local Poincar\'e invariance of the solutions,
we fix the gauge so that the metric is smooth outside the particles
and satisfies proper asymptotic conditions at space and time infinity.
We recover previous results for test bodies and find new ones for the
scattering of two dynamical particles in the massless limit.
\end{abstract}
\vfill
CERN-TH-6248/91\hfill\\
September 1991\hfill\\
\end{titlepage}
\pagenumbering{arabic}
\def\r{{(r)}}
\def\rp{{(r^\prime)}}
\def\1{{(1)}}
\def\2{{(2)}}
\def\p0{\varphi^0}
%

In a recent paper\footnote{Hereafter referred as (I).} \cite{CCV},
we have investigated the two-body classical scattering problem in
$2+1$ Gravity \cite{DJH}-\cite{Cappelli}, and we have provided an all-orders
solution for the scattering angle of two massless and spinless particles
in a properly defined center-of-mass frame.
The basic observation was that the Einstein theory corresponds to a gauge
fixing of the Chern-Simons theory of the Poincar\'e group \cite{Witten},
in which the metric is regular outside the particle trajectories, and
satisfies proper asymptotic conditions at space and time infinity.

The solution of the Einstein theory just mentioned was based on a class
of explicit N-particle singular solutions to the Chern-Simons theory.
Similar solutions have been found independently in the literature
\cite{Gupta,Grignani,Vaz}, sometimes including spin, but they are
not immediately useful for the Einstein theory, because they contain torsion
in the spinless limit.

In this note, we wish to generalize both the N-particle solution and
the classical scattering gauge fixing to the case of non-vanishing
intrinsic spin. Spinning particles can be thought of as localized
sources of torsion in the gravitational field equations.
Their introduction is natural in a first-order formalism and in a
Chern-Simons framework, in which the spin connection is coupled to the
angular momentum $J^a$.

However, the consequences of this treatment for the Einstein theory are not
yet fully understood, and may lead to conflict with causality
because of the apparence of time-like closed loops near to the torsion
source \cite{CTC}.
Here we find that the metric theory reconstructed from our solution
can also be interpreted as a torsionless Einstein theory, but with a
very singular energy-momentum tensor \cite{Hehl,Cartan}.
Our final result will be that the classical scattering occurs at the same
scattering angle, but with some time delay, with respect to its spinless
counterpart.

\bigskip
{\bf Equations of Motion and Einstein-Cartan Theory.}
\bigskip

To begin with, we shall write the Chern-Simons action with sources,
$\ S=S_{CS} + S_m $, as
\begin{eqnarray}
S
= & -& {1\over 2}\int d^3 x \ \ \epsilon^{\mu\nu\rho} \epsilon_{abc}
e^a_{\ \rho} \left( \partial_{[\mu} \omega_{\nu]} \ +\
\omega_{[\mu}\omega_{\nu]} \right)^{bc} \nonumber\\
&-&2 \sum_{(r)}\ \int d\tau \ \left[ \dot\xi^\mu\left( P_a \ e^a_{\ \mu} -
{1\over 2} J_a \epsilon^{abc}\omega_{\mu ,bc} \right)\right]_{(r)}\ \ ,
\label{CS}\end{eqnarray}
where $P^a_\r,J^a_\r$ are the internal momentum and angular momentum
of the $r$-th particle $(r=1,...,N)$ ,
and the dreibein $e^a_{\ \mu}$ and the spin connection
$\omega^a_{\mu\ b}$ are the components of the $ISO(2,1)$ gauge
connection \cite{Witten}.

The field equations corresponding to the action (\ref{CS}) are
given by
\begin{eqnarray}
\left( \partial_{[\mu} \omega_{\nu]} \ + \ \omega_{[\mu}
\omega_{\nu]}\right)^a_{\ b}
& = - &
\epsilon_{\mu\nu\lambda} \sum_{(r)} v^\lambda_{(r)} \epsilon^a_{\ bc}
P^{c}_{(r)} \delta^2 ( x - \xi_{(r)} (t) ) \ \ ,\label{omegaeq}\\
\left( \partial_{[\mu} e_{\nu]} \ + \ \omega_{[\mu}
e_{\nu]} \right)^a \ \
& = &
\epsilon_{\mu\nu\lambda} \sum_{(r)} v^\lambda_{(r)}
J^a_{(r)} \delta^2 ( x - \xi_{(r)} (t) )\ \ , \label{draieq}
\end{eqnarray}
where the momenta of the various particles are
$ p^{\mu}_{(r)} \ = \ m_{(r)} \frac{d\xi^{\mu}_{(r)}}{d\tau} \ \equiv
 \ m_{(r)}  {\dot\xi}^{\mu}_{(r)} ,\ \
v^\mu_{(r)}\equiv { {\dot\xi}^{\mu}_{(r)}/{\dot\xi}^0_{(r)}} $.
They show both an energy-momentum source and a torsion source.

The Bianchi identities for eqs.(\ref{omegaeq},\ref{draieq}) yield the
covariant conservation constraints for the particle variables,
which are
\begin{eqnarray}
\dot{P}^a_{(r)} &+& \dot\xi^\mu_{(r)} \omega^a_{\mu b}P^b_{(r)}=0 \ \ ,
\label{Peq}\\
{dJ^a_{(r)}\over d\tau} &+& \dot\xi^\mu_{(r)} \left(
\omega^a_{\mu b} J^b_{(r)} - \epsilon^a_{\ bc} P^c_{(r)} e^b_\mu
\right)=0 \ \ ,\label{Jeq}
\end{eqnarray}

As explained in I, we construct a metric theory of Einstein
type from the solution to eqs.(\ref{omegaeq}-\ref{Jeq}) by the
``soldering conditions''
\begin{eqnarray}
g_{\mu\nu}=e^a_{\ \mu}\eta_{ab} e^b_{\ \nu} \ \ ,\label{gmunu}\\
m_\r \ e^a_{\mu}(\xi_{(r)}) \ \dot\xi^{\mu}_{(r)} \ = \ P^a_{(r)},
\label{pmu}
\end{eqnarray}
which define the metric tensor $g_{\mu\nu}$ and the relation between
Lorentz and space-time momenta.

At this point, one should distinguish between the usual Levi-Civita
connection deduced from eq.(\ref{gmunu})
\beq
\Gamma_{\lambda,\mu\nu} \ = \ {1\over 2} \left(
\partial_{(\mu} g_{\nu)\lambda} -
\partial_{\lambda} g_{\mu\nu} \right)
\label{LCconn}\eeq
which is symmetric in $\mu$ and $\nu$, and the one reconstructed
from the spin connection $\omega_\mu$ in eqs.(\ref{omegaeq},\ref{draieq}),
i.e.,
\beq
\tilde\Gamma_{\lambda,\mu\nu} =e_{a\lambda}
\left( \eta^{ab}\partial_\mu +\omega_\mu^{ ab} \right) e_{b\nu}
\label{CSconn}\eeq
which is not symmetric, due to the existence of the torsion in
eq.(\ref{draieq}).
Notice that the parallel transport based on $\tilde\Gamma$ also preserves
the metric and thus the corresponding theory is of Einstein-Cartan type
\cite{Hehl}. In general, $\Gamma$ and $\tilde\Gamma$ are not
equal, but related by the identity
\beq
\Gamma_{\lambda,\mu\nu} - \tilde\Gamma_{\lambda,\mu\nu} =
{1\over 2} \left(
\tilde\Gamma_{\mu,[\nu\lambda]} + \tilde\Gamma_{\nu,[\mu\lambda]}
+\tilde\Gamma_{\lambda,[\nu\mu]} \right) \ \ ,
\label{Delta}\eeq
where the right hand side is given by eq.(\ref{draieq}) in terms of the
torsion tensor
\beq
T_{\lambda,\mu\nu} \equiv \tilde\Gamma_{\lambda,[\mu\nu]} =
\ e_{a\lambda} \epsilon_{\mu\nu\rho} \sum_{(r)} v^\rho_{(r)}
J^a_{(r)} \delta^2 ( x - \xi_{(r)} ) \ \ ,
\label{torsion}\eeq
which is non-vanishing at the localized spin sources.

A few consequences follow from eqs.(\ref{Delta}) and (\ref{torsion}).
First, in the spinless case we must have $\Gamma = \tilde\Gamma$,
and thus no torsion.
It follows that we cannot allow orbital angular momentum terms in
$J^a_\r$. Thus, we perform a Poincar\'e gauge choice by identifying
$J_\r$ with its intrinsic component $S_\r$, as follows,
\beq
J^a_\r = S^a_\r\equiv {\sigma_\r \over m_\r} \ P^a_\r \ \ ,
\label{spin}\eeq
instead of using the more general form
\beq
J^a_\r = {\sigma_\r \over m_\r} \ P^a_\r \ + \
\epsilon^a_{\ bc} \ b^b_\r P^c_\r.
\label{angular}\eeq
Note that in the three-dimensional Poincar\'e group, $S^a$ is proportional
to the momentum in terms of the spin scalar $\sigma_\r$.
Conversely, if we start from eq.(\ref{angular}) in the Chern-Simons theory, it
is always possible to perform a Poincar\'e gauge change of the form
\beq
\omega_\mu\to\omega_\mu,\ \ \ \ \ \ \ \
e_\mu\to e_\mu +(\partial_\mu + \omega_\mu)b(x), \ \ \ \ \
b(\xi_\r)=b_\r \ \ ,
\label{transf}\eeq
so as to eliminate the orbital part, leaving the Casimir
$(J_\r\cdot P_\r)=m_\r \sigma_\r $ invariant.
However, only the choice (\ref{spin}) defines, by eqs.(\ref{gmunu})
and (\ref{pmu}), a sensible Einstein theory having
$\Gamma=\tilde\Gamma$ for $\sigma_\r\to 0$, as explained
previously\footnote{
Thus we differ from refs.(\cite{Gupta,Grignani,Vaz}),
which do not distinguish between orbital and spin parts.
Another reason to restrict the torsion source is to obtain equivalence
between Poincar\'e gauge transformations and diffeomorphisms, as
discussed by Witten \cite{Witten}.
}.

The second observation is that, even if $\sigma_\r \neq 0$,
$\Gamma$ and $\tilde\Gamma$ coincide outside the particle trajectories,
because the torsion is localized (eq.(\ref{torsion})).
Thus, all consequences for parallel transporting a vector around the
sources with the $\Gamma$-connection can be obtained with
$\tilde\Gamma$ as well. Furthermore, it is easy to check that
$(\Gamma^\lambda_{\mu\nu} -\tilde\Gamma^\lambda_{\mu\nu})
\dot\xi^\mu_\r \dot\xi^\nu_\r $
vanishes on the particle sites, so that $\Gamma$ and $\tilde\Gamma$
are also equivalent for the particle motion.

Moreover, the Einstein equations for $\Gamma$ are, of course,
torsionless, but contain additional terms in the energy-momentum tensor
with respect to eq.(\ref{omegaeq}), because of the right hand side of
eq.(\ref{Delta}). Such terms from the spin sources have
$\delta^\prime$ -singularities, which can be
interpreted as localized sources of orbital angular momentum \cite{Cartan}.

\bigskip
{\bf N-Particle Solution.}
\bigskip

Let us now construct the solution to eqs.(\ref{omegaeq}-\ref{Jeq})
along the lines of I. To begin with, the particle motion is parametrized
in terms of three arbitrary functions $X^a(x^\mu)$ by the trajectory
equations
\beq\left\{\begin{array}{l l l}
X^2 (\xi^{\mu}_{(r)}) & = & B_{(r)} = {\rm const.}\\
X^1 (\xi^{\mu}_{(r)}) & = & V_{(r)} X^0 (\xi^{\mu}_{(r)})
\end{array}\right.\label{traj}\eeq
with the requirement
\beq
j_\r \ = \left| {\partial (X^1 \ - \ V_{(r)} X^0, X^2) \over
                  \partial ( x^1, x^2)} \right|_\r > 0 \ .
\label{jacob}\eeq

Here we notice that, in our spinning solution, some combinations of
the $X^a$'s (actually $X^0 -V_\r X^1$) will acquire a singularity,
at the particle sites. However, the combinations in eq.(\ref{traj})
turn out to be well defined, so that no problem arises.

The solutions to eqs.(\ref{omegaeq}-\ref{Jeq}) for $\omega$ and $e$
in the gauge (\ref{spin}) are characterized by constant $P^a_\r$
and are functions of $X^a(x)$. The spin connection has the same form
as in the case of spinless sources (I), i.e.,
\begin{eqnarray}
\omega_{\mu} & = & \sum_\r \omega_{\mu}^{(r)} \ \ ,\nonumber\\
\left( \omega_{\mu}^{(r)}\right)^a_{\ b} & = & \epsilon^a_{\ bc} \ P^c_{(r)}
\left( \partial_{\mu} X^2 \right) \delta ( X^2 -  B_{(r)} )
\Theta(V_{(r)} X^0 - X^1) \\
& \equiv & \epsilon^a_{\ bc} \ P^c_\r \ f_{\mu\r} \ \ ,\nonumber
\label{omegasol}\end{eqnarray}
for tails running on the left.

The dreibein contains instead an additional term related to the torsion
source, as follows,
\beq
e^a_{\ \mu} \ = \ \partial_{\mu} X^a \ + \ \sum_\r
\left( \omega_{\mu}^{(r)}\right)^a_{\ b} \ ( X^b - B^b_{(r)} )
- \ \sum_\r {\sigma_\r\over m_\r} P^a_\r f_{\mu\r}\ \ ,
\label{esol}\eeq
where we have explicitly used the intrinsic spin identification in
eq.(\ref{spin}).

We first need to check eq.(\ref{draieq}), which is actually linear
in $e^a_{\ \mu}$. Therefore, the additional spin term in eq.(\ref{esol})
directly gives rise to the torsion source, as in the equation (\ref{omegaeq})
for the spin connection, because the quadratic terms vanish again.
For $\sigma\to 0$, eq.(\ref{esol}) reduces, of course, to the (non-trivial)
solution of the torsionless case, discussed in I, which already provides
the relevant orbital angular momentum terms due to the translational
parameters $B_\r$.

Next, we consider the covariant conservation constraints.
Equation (\ref{Peq})
is trivially verified by a constant $P^a$, because $(1+\omega_\mu)$
leaves it invariant (I). Equation (\ref{Jeq}) is also satisfied
for constant $J^a$, parallel to $P^a$, because the last term
drops out, owing to eq.(\ref{pmu}).
The consistency equation (\ref{pmu}) is less trivial, due to the
extra singularities introduced by torsion. In fact, by using the
explicit form of $e^a_{\ \mu}$ in eq.(\ref{esol}) we obtain, for
the $r$-th particle,
\beq
P^a_\r \ = \ \left[
m_\r {d\over d\tau} X^a(x) - \ \sigma_\r P^a_\r \
\Theta \left( V_\r X^0(x) -X^1(x) \right) \
{d\over d\tau} \Theta\left( X^2(x) - B_\r \right)
\right]_{x=\xi_\r(\tau)} \ \ .
\label{disco}\eeq

Since $P^a$ is constant by hypothesis, eq.(\ref{disco}) shows that
$X^a$ is ill-defined at the particle site by a quantity proportional
to $\sigma P^a/m$, if the limit $x\to\xi$ is performed on the tail side.
Fortunately, the ill-defined term cancels for the quantities
$(X^1-V_\r X^0)$ and $(X^2-B_\r)$ (which are still defined and vanish on
the trajectory), provided we set
\beq
V_r ={P^1_\r \over P^0_\r} \ .
\label{velo}\eeq

On the other hand, the combination $(X^0 - V_\r X^1)$ possesses a
discontinuity on the tail given by
\beq
\Delta \left( \gamma (V_\r)(X^0 -V_r X^1)\right) = - \sigma_\r \ \ ,
\label{disco2}\eeq
which corresponds to the well-known time jump \cite{DJH}.
Such pathology of the $X$ variable will disappear in the regular $x$
variable to be  discussed below.

The time jump feature is confirmed from the analysis of the matching
conditions for geodesics crossing the tail. The Einstein spacetime is
defined by eqs.(\ref{gmunu},\ref{LCconn}), but eq.(\ref{Delta})
implies that the geodesics $x^\mu(\tau)$ going across the tail,
but not hitting the particle, are still determined
by the Chern-Simons connection $\tilde\Gamma$.
Thus they satisfy the first-order equation
\beq \frac{d}{d\tau} ( e^a_{\ \nu}{\dot{x}^{\nu}} ) + {\dot{x}}^{\mu}
\omega^a_{\mu\ b} ( e^b_{\ \nu} {\dot{x}}^{\nu} ) \ = \ 0 \ .
\label{geo}\eeq
By integrating twice as in I, we obtain
for a point just above $(X_+)$ and just below $(X_-)$ the tail
(running on the left of the particle),
\begin{eqnarray}
\left( X_- - X(\xi_{(r)}) \right)^a &=& L^a_{\r\ b}
\left( X_+ -{\sigma_\r\over m_\r} P_\r - X(\xi_{(r)}) \right)^b \ \ ,
\label{geodisc}\\
L_{(r)} \ &=& \ e^{- P_\r^a {\cal J}_a }
 \ \ \ \ \ ({\rm r-th \ \ tail \ } )
\label{Lr}\end{eqnarray}
(using the representation $({\cal J}_a)^b_{\ c}= \epsilon^b_{\ ac}$).
This relation shows the well-known matching condition of Deser, Jackiw
and 't Hooft \cite{DJH}, including its space-time jump.

Let us finally note that our gauge choice $J_\r = ({\sigma\over m}P)_\r$
is consistent in the case of non-parallel velocities, i.e. when
the particle $(1)$ crosses the tail of particle $(2)$.
A  rigorous discussion of this point needs the analysis of the Poincar\'e
holonomies (see below). However, the final result is a simple consequence
of the conservation constraints (\ref{Peq},\ref{Jeq}), because, by
eq.(\ref{pmu}), the last term of eq.(\ref{Jeq}) vanishes identically.
Thus, at the tail crossing, the $\omega_2$-term rotates both $P_\1$ and $J_\1$
in the same way, so that they are still proportional after crossing.

\bigskip
{\bf Holonomies and Particle Exchanges.}
\bigskip

{}From the Chern-Simons point of view, only topological quantities are
observable, and come from the invariants of Poincar\'e holonomies around
the particles.
The single-particle holonomies, already computed in I, are given by
\beq
U_\r(x_\r,x_\r)=
P \exp \left(-\oint_{\xi_\r +\epsilon}
     \omega\cdot{\cal J} + e\cdot {\cal P}\right) =
\left( \begin{array}{l l}
e^{-P_\r\cdot{\cal J}} &  - {\sigma_\r\over m_\r}P_\r \\
                     0 & 1 \end{array} \right) \ \ ,
\label{Ur}\end{equation}
where we used the $4\times 4$ representation for the generators
${\cal J}_a, {\cal P}_a$ of the Poincar\'e group
and the gauge choice (\ref{spin}).
By a gauge trasformation at the basepoint $g=g(x_\r)$,
$U_\r \to gU_\r g^{-1}$, and its invarants are given by
\begin{eqnarray}
m_\r & \ \leftrightarrow \ \ &
{\rm Tr}\left(e^{-P_\r\cdot{\cal J}}\right) = 1+2\cos m_\r \ \ ,\label{minv}\\
\sigma_\r & \ \leftrightarrow \ \ &
J_\r\cdot P_\r = \sigma_\r m_\r \ .\label{sinv}
\end{eqnarray}
Thus the $U_\r$ invariants correspond to the constants
of motion of the r-th particle, i.e. its mass and spin.

The two-particle system is characterized by the O-loop, i.e. by the
holonomy encircling once (counterclockwise) both particles.
Taking into account the translation between the two particles of
impact parameters $B_\1$ and $B_\2$ and using the result (\ref{Ur})
for each particle, we obtain
\beq
U_O = P \exp \left(-\oint_{1+2}
     \omega\cdot{\cal J} + e\cdot {\cal P}\right) =
\left( \begin{array}{c c}
e^{w\cdot{\cal J}} & q \\
                     0 & 1 \end{array} \right) \ \ ,
\label{Oloop}\end{equation}
with invariants
\beq\begin{array}{r l}
\cos{{\cal M}\over 2} = & \cos {\sqrt{w^2}\over 2}= \
c_\1 c_\2 - s_\1 s_\2 {P_\1 \cdot P_\2\over m_\1 m_\2} \ \ ,\\
{\cal S} =& {w\cdot q\over {\cal M}} =
\left( \sin{{\cal M}\over 2}\right)^{-1} \left[
2 s_\1 s_\2 \epsilon_{abc} (B_\1 -B_\2)^a {P_\1^b P_\2^c \over m_\1 m_\2} +
\right. \\
\ \ & \ \ \ \ \ \ \ \left.
\sigma_\1 \left(c_\2 s_\1 - c_\1 s_\2 {P_\1 \cdot P_\2\over m_\1 m_\2}\right)
+ \sigma_\2 \left(c_\1 s_\2 - c_\2 s_\1 {P_\1 \cdot P_\2\over m_\1 m_\2}\right)
\right]\ \ ,
\end{array}\label{oinv}\eeq
where $c_{(i)}= \cos{m_{(i)}\over 2}, s_{(i)}=\sin{m_{(i)}\over 2}$.
In the special relativity limit $(m_{(i)} \equiv 8\pi G m_{(i)} \ll 1)$,
eq.(\ref{oinv}) reduces to
\begin{eqnarray}
{\cal M}^2 &=& ( P_\1 +P_\2 )^a \ \eta_{ab}\ ( P_\1 + P_\2 )^b \ \ ,\nonumber\\
{\cal S} &=& {(P_\1 +P_\2)^a \over{\cal M}}\ \eta_{ab} \left[
S^a_\1 +S^a_\2 + \epsilon^a_{\ bc} (B^b_\1 P^c_\1 + B^b_\2 P^c_\2) \right]
\ \ ,\label{smallm}\end{eqnarray}
consistent with the usual interpretation of the total mass and total
angular momentum.

As discussed in I, the time evolution can lead to the collisions of
two particles, which is represented by an ``exchange'' operation
$\sigma_{12}$, corresponding to particle $2$ crossing the tail of
particle $1$ (and vice versa for $\sigma_{21}$).
This exchange operation can be studied as in I by means of
the non-abelian Stokes theorem for the Poincar\'e holonomies, and
has the presentation
\beq
\sigma_{12}\ :\ \ \left\{ \begin{array}{l}
      U_\1 \rightarrow  U_\1 \\
      U_\2 \rightarrow  U_\1  U_\2  U_\1^{-1}
\end{array} \right.
\label{braiding}\eeq
where the $U_\r$'s are the single-particle holonomies (\ref{Ur})
referred to a common basepoint, with a prescribed tail orientation,
e.g. to the left of
all particles\footnote{The monodromy operation, given by
$\sigma_{12}\sigma_{21}$, is instead gauge invariant.}.

By taking into account the spin part in the $U_\r$'s, one obtains from
(\ref{braiding}) the following transformation properties
\beq
\sigma_{12}\ : \ \left\{ \begin{array}{l l l}
B_\1 \rightarrow B_\1 &,& B_\2 \rightarrow B_\1 + L_\1 ( B_\2 - B_\1 ) - S_\1\\
P_\1 \rightarrow P_\1 &,& P_\2 \rightarrow L_\1  P_\2  \\
S_\1 \rightarrow S_\1 &,& S_\2 \rightarrow L_\1 S_\2
\end{array} \right.
\label{twodisc}\eeq
The first two equations are indeed equal to the geodetic limit for particle
$(2)$ discussed before (eq.(\ref{geodisc})), and the third one implies
that $S_\2$ stays parallel to $P_\2$, as anticipated. Therefore, our
gauge choice eq.(\ref{spin}) is consistent with the most general motion
of the particles.

\bigskip
{\bf Smooth Spacetimes and Scattering Properties.}
\bigskip

Let us finally discuss the spacetime corresponding to our solution
in the Einstein theory, as defined by eqs.(\ref{gmunu}) and
(\ref{pmu}).
On physical grounds, we require the metric (\ref{gmunu})
to be (i) regular outside the particle trajectories, as explained in I.
Furthermore, we
impose (ii) asymptotic conditions defining single-particle and two-particle
states at large negative times, and (iii) forbid rotating frames at spatial
infinity.

The above conditions can be enforced by using the residual gauge freedom
of our solutions, given by Poincar\'e transformations leaving
$\{ P^a, S^a={\sigma\over m}P^a\}$ invariant, i.e. by (pseudo)-rotations
around $P^a$ and traslations along $P^a$.
We can thus look for gauge fixing on the coordinates
$X^A=\overline{X}^A, (A=0,1,2,3)$, defined in the $4\times 4$
representation by
\beq
\overline{X}^A \ = \ {({\cal{T}} (x))}^A_{\ B} \ x^B,
\eeq
where the local Poincar\'e transformation ${\cal T}$ is chosen so as
to build a solution of the matching conditions around  each particle
site. Thus, for each $\r$, we require
\beq
{\cal{T}} \simeq {\cal{T}}_{(r)} (x) S_{(r)} (x) , \ \ \ \ \ \
{\rm for} \ \ \ x^\mu \simeq \xi^\mu_{(r)} (\tau),
\label{factoriz}\eeq
where $S_\r $ is regular for $x^\mu \to \xi^\mu_\r$.
For a particle passing through the origin,
${\cal{T}}$ performs the change of variables
\beq
\overline{X}^a \ = \Lambda^a_{\ b} x^b \ + \
{\sigma_\r\over m_\r} P^a_\r {{\p0}_\r (x)\over 2\pi }
\ \ ,\ \ \ \ \
\Lambda \equiv \exp\left({\p0_\r(x)\over 2\pi} P_\r\cdot{\cal J}\right)\ \ ,
\label{Lambda}\eeq
\beq
\tan \p0_\r = \frac{1}{\gamma_\r} \frac{y}{x- V_\r t} \ \ ,
\ \ \ \ \ P^{\mu}_\r = m ( \gamma, V \gamma ,0) \vert_\r \ \ ,
\label{tanpi}\eeq
where $\p0_r$ is the azimuthal angle in the r-th particle rest frame.
The trasformation (\ref{Lambda}) differs from the one given in I only for the
spin-dependent translation, which reproduces the time jump in $X$-variables
as $\p0_\r$ varies from $-\pi$ to $+\pi$, eq.(\ref{geodisc}),
such that the $x$-variables is
continuous everywhere, and fulfils the first requirement above.

If we have only one particle, it is easy to provide the explicit form
of the dreibein resulting from the gauge fixing (\ref{Lambda}) of
eq.(\ref{esol}), i.e.,
\begin{eqnarray}
e^a_{\ \mu} &=& \left(\partial_{\mu} + \omega_{\mu} \right)^a_{\ b}
\overline{X}^b - {\sigma\over m}P^a f_\mu \nonumber\\
&=& \Lambda^a_{\ b} \left[ \delta^b_\mu +\left(
\left( P\cdot{\cal J}\right)^b_{\ c}x^c +{\sigma \over m}P^b \right)
{1\over 2\pi}\partial_\mu\p0 \right].
\label{singlepart}\end{eqnarray}
Notice that the explicit singular parts of $\omega_\mu$ occurring
in $e^a_{\ \mu}$ have been cancelled by the singular derivative terms of the
$\overline{X}^a$-variable in (\ref{Lambda}). More explicitly we obtain
\begin{eqnarray}
e^a_{\ \mu} &=& \Lambda^a_{\ b}
\left( \delta^b_{\mu} + {m\over 2\pi}{N^b N_{\mu} \over |N^2|}
+{\sigma\over 2\pi}{P^b\over m}{N_\mu\over |N^2|} \right), \\
N^a(P) &=& \left( {P\over m}\cdot{\cal J}\right)^a_{\ b} x^b \ \
= \gamma\left( Vy, y, -x+Vt\right) \ \ ,\nonumber
\label{esmooth}\end{eqnarray}
and, therefore, from eq.(\ref{gmunu})
\beq
g_{\mu\nu} = \eta_{\mu\nu} +
{N_{\mu} N_{\nu} \over |N^2|}( 1 - {\alpha}^2)+
{N_{\mu} N_{\nu} \over N^4}\left({\sigma\over 2\pi}\right)^2+
{\sigma\over 2\pi |N^2|}\left(N_\mu {p_\nu\over m}+N_\nu {p_\mu\over m}\right)
\ \ ,\label{gsmooth}\eeq
where $\alpha =1-{m\over 2\pi}$.
This metric reduces to the well-known one of the spinning cone
\cite{DJH,Spincone} in the static case
$\ (\p0\equiv\varphi)$ , i.e.,
\beq
ds^2 = \left( dt+ {\sigma\over 2\pi}d \varphi\right)^2 -
dr^2 -  r^2 {\alpha}^2 d \varphi^2 \ .
\label{scone}\eeq

We can also study the massless limit in eq.(\ref{gsmooth}), by setting
\beq
S^a=\lambda P^a \ \ \ \ \ \ \ \ \ \ ( m=0 ),
\label{m0spin}\eeq
i.e. $\sigma=m\lambda$, where the (Poincar\'e) scalar $\lambda$ is
kept fixed, while the three-dimensional ``helicity''
$\vec{S}\cdot\vec{P}/ |\vec{P}|=\lambda E$
transforms as the zeroth component of a vector.
In this limit, we obtain the ``spinning'' Aichelburg-Sexl metric
\cite{AS}
\beq
ds^2 \ = 2 dudv - (dy)^2 + \sqrt{2} E \delta(u) \left(
|y| + \lambda \ {\rm sign}(y)\right) (du)^2 \ \ ,
\label{gAS}\eeq
where $u = \frac{1}{\sqrt{2}}(t-x) , \ v = \frac{1}{\sqrt{2}} (t+x)\ \ $
are light-cone variables.

Next, we consider the test particle scattering in the one-particle metric
given above. From eqs.(\ref{esmooth}) and (\ref{gAS}), one sees
that the spin modifications do not affect the scattering angle,
because they are negligible at large distances; they do, however,
give a time delay with respect to the spinless case of I.
This can be seen more explicitly from the geodesic trajectory
in the smooth metric produced by a particle of momentum $P^a$ and
spin $\sigma$, namely
\beq
x^a (\tau) \ = - {\sigma\over m} P^a {\phi^0(\tau)\over 2\pi\alpha}
+ \exp\left(-\frac{\phi^0(\tau)}{2\pi\alpha}\right)^a_{\ b}
\ ( U^b \ \tau \ + \ B^b ) \ \ ,
\label{xtraj}\eeq
where $\phi^0(\tau)\equiv\alpha\p0$ is the azimuthal variable of the
geodesic in Minkowskian $X$-variables $ X(\tau) =U\tau + B $
and in the particle rest frame.
During the evolution, $\phi^0$ changes from $\phi^0=0$ to $\phi^0=\pi$,
so that the classical ``scattering matrix'' has a Lorentz transformation
and a translation given by
\beq
L(P)^{\pm {1\over 2\alpha}} \ \ \ , \ \ \ \
\Delta x^\mu= \mp {\sigma\over 2\alpha} {p^\mu\over m}\ \ ,
\label{geoscatt}\eeq
for geodesics running above (below) the spinning particle.
In the static limit, the translation reduces to the geodetic ``time delay''
alread found in ref.\cite{Spincone}.

Finally, the scattering problem for two particles is set up along the same
lines as in I, the only difference being that one has to keep track of the
space-time jump associated with the $\Lambda$-transformation.
For the massless case, by imposing  the absence of impulsive rotations
and translations at infinity, a condition follows for the collision
at $t=0$ of two Aichelburg-Sexl metrics.
This yields the scattering parameters \cite{CCVV}
\begin{eqnarray}
p_f &=& R(\theta) p_i \ \ ,\ \ \ \tan {\theta\over 2}={\sqrt{s}\over 4} \ \ ,
\nonumber\\
\Delta x^\mu &=& - \left( \frac{\lambda_1 p_1 + \lambda_2 p_2}{2}
\epsilon (b) + \frac{|b|}{4} ( p_1 + p_2 ) \right) + O(s) \ \ ,
\label{twoscatt}\end{eqnarray}
in the center-of-mass defined in I.
Equation (\ref{twoscatt}) shows the space-time jump for the scattering
of two dynamical particles, whose final (initial) momenta are
denoted by $p_f (p_i)$, \hbox{$p_\1=(E,E,0)$},$\ \  p_\2=(E,-E,0)$,
$\ s=(p_\1+p_\2)^2 \ $ and $\epsilon = +1 (-1)$ corresponds to impact
parameters of positive (negative) angular momenta.

\vfill\eject
\def\NP{Nucl. Phys.\ }
\def\AP{Ann. Phys. (NY)\ }


\begin{thebibliography}{99}
\bibitem{CCV}
A. Cappelli, M.Ciafaloni and P.Valtancoli, {\it Classical Scattering
in 2 + 1 Gravity with N Point Sources}, preprint CERN-TH-6093/91,
May 1991.
%
\bibitem{DJH}
S. Deser, R. Jackiw and G. 't Hooft, \AP {\bf 152} (1984) 220.
%
\bibitem{H}
G. 't Hooft, Comm. Math. Phys. {\bf 117} (1988) 685.
%
\bibitem{DJ}
S. Deser and R. Jackiw, Comm. Math. Phys. {\bf 118} (1988) 495.
%
\bibitem{Spincone}
P. Sousa Gerbert and R. Jackiw, Comm. Math. Phys. {\bf 124} (1989) 229.
%
\bibitem{Witten}
E. Witten, Nucl. Phys. {\bf B311} (1988-89) 46, {\bf B323} (1989) 113.
%
\bibitem{Carlip}
S. Carlip, Nucl. Phys. {\bf B324} (1989) 106.
%
\bibitem{Gerbert}
P. Sousa Gerbert, Nucl. Phys. {\bf B346} (1990) 440.
%
\bibitem{LWitten}
K.Koehler, F.Mansouri, C.Vaz and L.Witten, \NP {\bf B348} (1991) 373.
%
\bibitem{Cappelli}
A. Cappelli, talk to the {\it Workshop on Random Surfaces and 2-D
Quantum Gravity,} June 1991, Barcelona, preprint CERN-TH-6249/91,
to appear in Nucl. Phys. B (Proc. Suppl.).
%
\bibitem{Gupta}
K. Gupta and A. Stern, {\it Spin and Statistics
in $2+1$ Gravity}, preprint Syracuse SU-4228-463 (1991).
%
\bibitem{Grignani}
G.Grignani and G.Nardelli, {\it Gravity in 2 + 1 Dimensions Coupled
to Point-Like Sources: A Flat Chern-Simons Gauge Theory Equivalent
to Einstein}, preprint MIT CPT-1953 (1991).
%
\bibitem{Vaz}
C.Vaz and L.Witten, {\it On the Multiparticle Phase Space in
Chern-Simons Gravity}, preprint UCTP-103/91.
%
\bibitem{CTC}
G.Clement, I.J.Theor.Phys. {\bf 24} (1985) 267;\\
S.Deser, R.Jackiw and G. 't Hooft, {\it Physical Cosmic Strings do not
Generate Closed Time-like Curves}, preprint MIT CPT-2011 (1991).
%
\bibitem{Hehl}
F.W.Hehl, P. Heyde and G.D.Kerlick, Rev. Mod. Phys. {\bf 48} (1976) 393.
%
\bibitem{Cartan}
W.Arkuszewski, W.Kopczynski and V.N.Ponomariev, Ann. Inst.H.Poincar\'e
{\bf A 21} (1974) 89.
%
\bibitem{AS}
P.C. Aichelburg and R.U. Sexl, Gen. Rel. Grav. {\bf 2} (1971) 303; \\
T. Dray and G. 't Hooft, Nucl. Phys. {\bf B253}  (1985) 173.
%
\bibitem{CCVV} A.Cappelli, M.Ciafaloni and P.Valtancoli, to appear.
%
\end{thebibliography}
\end{document}